\documentclass[conference]{IEEEtran}
\IEEEoverridecommandlockouts
\usepackage{cite}
\usepackage{amsmath,amssymb,amsfonts}
\usepackage{algorithmic}
\usepackage{graphicx}
\usepackage{textcomp}
\usepackage[dvipsnames]{xcolor}
\usepackage{listings} 
\usepackage{multicol} 
\usepackage{multirow} 
\usepackage{caption} 
\usepackage[utf8]{inputenc} 
\usepackage{colortbl} 
\usepackage{hhline}
\usepackage[hidelinks]{hyperref}
\usepackage{array}
\usepackage{ragged2e}
\usepackage{soul}
\usepackage{xcolor}
\usepackage{color}
\usepackage{tabularray}
\sethlcolor{lightgray} 
\setul{0.5ex}{0.3ex} 

\let\othelstnumber=\thelstnumber
\def\createlinenumber#1#2{
    \edef\thelstnumber{%
        \unexpanded{%
            \ifnum#1=\value{lstnumber}\relax
              #2%
            \else}%
        \expandafter\unexpanded\expandafter{\thelstnumber\othelstnumber\fi}%
    }
    \ifx\othelstnumber=\relax\else
      \let\othelstnumber\relax
    \fi
}
\bgroup

\def\BibTeX{{\rm B\kern-.05em{\sc i\kern-.025em b}\kern-.08em
    T\kern-.1667em\lower.7ex\hbox{E}\kern-.125emX}}
\begin{document}

\title{RISC-V R-Extension: Advancing Efficiency with Rented-Pipeline for Edge DNN Processing}
\author{\IEEEauthorblockN{Won Hyeok Kim}
\IEEEauthorblockA{\textit{Department of Artificial Intelligence}\\
\textit{Sungkyunkwan University}\\
Suwon, Korea\\
2080fresh@skku.edu}
\and
\IEEEauthorblockN{Hyeong Jin Kim}
\IEEEauthorblockA{\textit{Department of Semiconductor}\\\textit{Convergence Engineering}\\
\textit{Sungkyunkwan University}\\
Suwon, Korea\\
wsx0409@g.skku.edu}
\and
\IEEEauthorblockN{Tae Hee Han}
\IEEEauthorblockA{\textit{Department of Semiconductor}\\\textit{Systems Engineering}\\
\textit{Sungkyunkwan University}\\
Suwon, Korea\\
than@skku.edu}}

\maketitle

\begin{abstract}
The proliferation of edge devices necessitates efficient computational architectures for lightweight tasks, particularly deep neural network (DNN) inference. Traditional NPUs, though effective for such operations, face challenges in power, cost, and area when integrated into lightweight edge devices. The RISC-V architecture, known for its modularity and open-source nature, offers a viable alternative. This paper introduces the RISC-V R-extension, a novel approach to enhancing DNN process efficiency on edge devices. The extension features rented-pipeline stages and architectural pipeline registers (APR), which optimize critical operation execution, thereby reducing latency and memory access frequency. Furthermore, this extension includes new custom instructions to support these architectural improvements. Through comprehensive analysis, this study demonstrates the boost of R-extension in edge device processing, setting the stage for more responsive and intelligent edge applications.
\end{abstract}


\section{Introduction}
In the rapidly evolving landscape of computing, the efficiency of processing units is paramount, especially for small-edge devices like home appliances and IoT devices. These devices require deep neural network (DNN) inference workloads that adhere to strict power and space limitations. The RISC-V architecture, known for its open-source and modular nature, is a promising candidate for these challenges.

Central to edge device functionality is the processing of DNNs. Recently, NPUs and TPUs have been the mainstay for such tasks in smartphone-size devices. These specialized units offer optimized performance for neural network computations but are often impractical for integration into smaller, more resource-constrained edge devices due to their size, power, and cost. Consequently, there is a growing need for efficient CPU-based solutions, especially for devices that cannot accommodate separate neural processing hardware. 

Previous advancements in RISC-V for DNN processing include F-extension for floating-point (FP) processing and multiply-accumulate (MAC) instructions. However, this research introduces R-extension, which improves upon these earlier developments. With its unique rented-pipeline mechanism and use of architectural pipeline registers (APR), the R-extension is tailored to enhance the processing of DNNs in edge devices. The main contributions of R-extension are as follows:

\begin{itemize}
\item Rented-pipeline: Rent a memory (MEM) stage on the 5-stage pipeline when fetching proposed instructions. Renting a MEM stage allows the processor to consume two stages during execution without increasing the number of stages or critical path delay. 
\item APR (Architectural Pipeline Register): Located on the MEM/write-back (WB) pipeline register, accumulated results are constantly updated in the APR. By using APR, the processor does not need to use memory for read and write during MAC operations. When accumulation ends, the processor writes back APR data into the destination register and resets APR.
\item Custom instructions: This paper represents two new instructions, \textit{rfmac.s} and \textit{rfsmac.s}, to support the previous two contributions. \textit{rfmac.s} stands for single-precision FP MAC operation on R-extension, and \textit{rfsmac.s} stands for store result of single-precision FP MAC operation working on R-extension.
\item  Versatility: The proposed microarchitecture facilitates the addition of further instructions for diverse accumulation operations, such as integer MAC operation and difference accumulator \cite{b12}, enhancing its alterability. The potential integration into vector architectures also paves the way for increased parallelism, offering prospects for system speed enhancements. 
\end{itemize}

The performance metrics of the R-extension are noteworthy. Compared to the F-extension's FP multiplication approach (RV64F), our R-extension (RV64R) achieves an increment of 29\% in instructions-per-cycle (IPC) and a 34\% decrement in the number of memory accesses. Furthermore, compared to the baseline that solely added a MAC instruction without altering the pipeline, RV64R exhibits 15\% IPC and 22\% memory access improvement with a negligible increase in area.

Advancements in R-extension can improve edge device processing for lightweight AI tasks, achieving remarkable efficiency and adaptability. This paper will delve into the details of the R-extension and its comparative advantages.

\begin{figure*}[bth]
    \begin{multicols}{2}
    \createlinenumber{11}{.} \createlinenumber{12}{.}
    \begin{minipage}{.48\textwidth}
     \begin{lstlisting}[language=C, gobble=5, numbers=none]
     // (*@\textcolor{gray}{H = H$_{\text{in}}$ - H$_{\text{fil}}$ + 1}@*)
     // (*@\textcolor{gray}{W = W$_{\text{in}}$ - W$_{\text{fil}}$ + 1}@*)  
     for (i = 0; i < M; i++)
      for (j = 0; j < H; j += S)
       for (k = 0; k < W; k += S)
        for (l = 0; l < C; l++)  
         for (m = 0; m < H(*@$_{\text{fil}}$@*); m++)
          for (n = 0; n < W(*@$_{\text{fil}}$@*); n++)
           Output[i][j/S][k/S](*@\textcolor{OrangeRed}{+=}@*)
           Input[l][j+m][k+n](*@\textcolor{OrangeRed}{*}@*)Filter[i][l][m][n]);
    
                          (*@\textrm{(a-1)}@*)
    \end{lstlisting}
     \end{minipage}
    

     \begin{minipage}{.47\textwidth}
     \begin{lstlisting}[language=C, gobble=5, numbers=none]
     000 <convolve>:
     000: j	 182        (*@\hl{0ca: fmul.s	fa5,fa3,fa5}@*)
     006: j   16a        (*@\hl{0e4: fadd.s	fa5,fa4,fa5}@*)
     00c: j	 152        (*@\hl{0fc: fsw \ \ \ fa5,0(a5) \ \ }@*)
     012: j	 13a        114: bge	a5,a4,020
     018: j	 122        12c: bge	a5,a4,01a
     01e: j	 10a        144: bge	a5,a4,014
     (*@\hl{04a: flw fa4,0(a5)}@*)  15c: bge	  a5,a4,00e 
     (*@\hl{08e: flw fa3,0(a5)}@*)  174: bge	  a5,a4,008
     (*@\hl{0c6: flw fa5,0(a5)}@*)  18c: bge	  a5,a4,002
     
                          (*@\textrm{(a-2)}@*)
     \end{lstlisting}
     \end{minipage}
     \end{multicols}

     \begin{multicols}{2}
     \begin{minipage}{.48\textwidth}
     \begin{lstlisting}[language=C, gobble=5, numbers=none]
     for (i = 0; i < M; i++)
      for (j = 0; j < H; j += S)
       for (k = 0; k < W; k += S)
        for (l = 0; l < C; l++)
         for (m = 0; m < H(*@$_{\text{fil}}$@*); m++)
          for (n = 0; n < W(*@$_{\text{fil}}$@*); n++)
           fmac_s(Output[i][j/S][k/S],
           Input[l][j+m][k+n],Filter[i][l][m][n]);
    
    
    
                          (*@\textrm{(b-1)}@*)
     \end{lstlisting}
     \end{minipage}
    
     \begin{minipage}{.47\textwidth}
    
     \begin{lstlisting}[language=C, gobble=5, numbers=none]
     000 <convolve>:
     000: j	 154        (*@\hl{0ca: fmac.s	fa5,fa4,fa3}@*)
     006: j	 13c        (*@\hl{0ce: fsw \ \ \ fa5,0(a3) \ \ }@*)
     00c: j	 124        0e6: bge	a5,a4,020
     012: j	 10c        0fe: bge	a5,a4,01a
     018: j	 0f4        116: bge	a5,a4,014
     01e: j	 0dc        12e: bge	a5,a4,00e
     (*@\hl{08a: flw fa4,0(a5)}@*)  146: bge	  a5,a4,008
     (*@\hl{0c2: flw fa3,0(a5)}@*)  15e: bge	  a5,a4,002
     (*@\hl{0c6: flw fa5,0(a3)}@*)
    
                          (*@\textrm{(b-2)}@*)
    \end{lstlisting}
     \end{minipage}
     \end{multicols}

     \begin{multicols}{2}
     \begin{minipage}{.48\textwidth}
     \begin{lstlisting}[language=C, gobble=5, numbers=none]
     for (i = 0; i < M; i++)
      for (j = 0; j < H; j += S)
       for (k = 0; k < W; k += S)
        for (l = 0; l < C; l++)
         for (m = 0; m < H(*@$_{\text{fil}}$@*); m++)
          for (n = 0; n < W(*@$_{\text{fil}}$@*); n++)
           rfmac_s(Input[l][j+m][k+n],
           Filter[i][l][m][n]);
        rfsmac_s(Output[i][j/S][k/S]);
    
          
                          (*@\textrm{(c-1)}@*)
     \end{lstlisting}
     \end{minipage}
    
     \begin{minipage}{.47\textwidth}
     \begin{lstlisting}[language=C, gobble=5, numbers=none]
     000 <convolve>:
     000: j	 154        (*@\hl{09c: rfmac.s \ fa5,fa4 \ \ }@*)
     006: j	 13c        0b4: bge	  a5,a4,020
     00c: j   124	    0cc: bge	  a5,a4,01a
     012: j	 0da        0e4: bge	  a5,a4,014
     018: j	 0c2        0fe: rfsmac.s fa5
     01e: j	 0aa        116: fsw	  fa5,0(a5)
     (*@\hl{060: flw fa5,0(a5)}@*)  12e: bge	    a5,a4,00e
     (*@\hl{098: flw fa4,0(a5)}@*)  146: bge	    a5,a4,008
                         15e: bge	  a5,a4,002
    
                          (*@\textrm{(c-2)}@*)
    \end{lstlisting}
    \end{minipage}
     \end{multicols}
     \caption{(a) RV64F, (b) Baseline, (c) RV64R. Left: Convolution code, H$_{\text{in}}$: Input Height, W$_{\text{in}}$: Input Width, M: Number of Filter, C: Channel, H$_{\text{fil}}$: Filter Height, W$_{\text{fil}}$: Filter Width. Right: Assembly language after compile. Highlighted parts are the main instructions in the most inner of all loops.}
     \label{fig:code} 
\end{figure*}  

\section{Background}

\subsection{Advanced Pipeline}
The challenge in processing DNN inference on CPUs, particularly in lightweight devices, is managing frequent memory usage, which is essential to maintaining process capability within the power constraints of small-scale devices. In the context of RISC-V implementations for such devices, complex pipelining techniques like those used in superscalar processors, generally optimized for high performance CPUs, are unsuitable due to their high power consumption and area costs.

MAC operations, characterized by repetitive multiply and add sequences, are distinct from complex operations requiring diversified pipelines. These diversified pipelines, which are effective for handling extended latency operations, are not optimized for the continuous accumulation required in MAC operations and also involve higher area costs and power consumption because of more pipeline stages, which are critical constraints in small devices \cite{b10}. In contrast, using rented-pipelining, the R-extension for RISC-V offers an accelerated solution for compact devices using the execution (EX) stage on multiplication and the MEM stage on accumulation. This approach is bespoken for MAC operations, which are fundamental in DNN inferences.

In MAC operations within assembly sequences shown at \autoref{fig:code}(a-2), read-after-write (RAW) hazards occur when the processor attempts to accumulate new multiplication results with data not yet updated from a previous WB stage. Conventional data forwarding methods, which preempt RAW hazards by bypassing register write and read cycles, may not suffice for MAC scenarios where accumulated data is pending in the write-back stage and not immediately available for forwarding.

\begin{figure}[t!]
    \centering
    \includegraphics[width=1\linewidth]{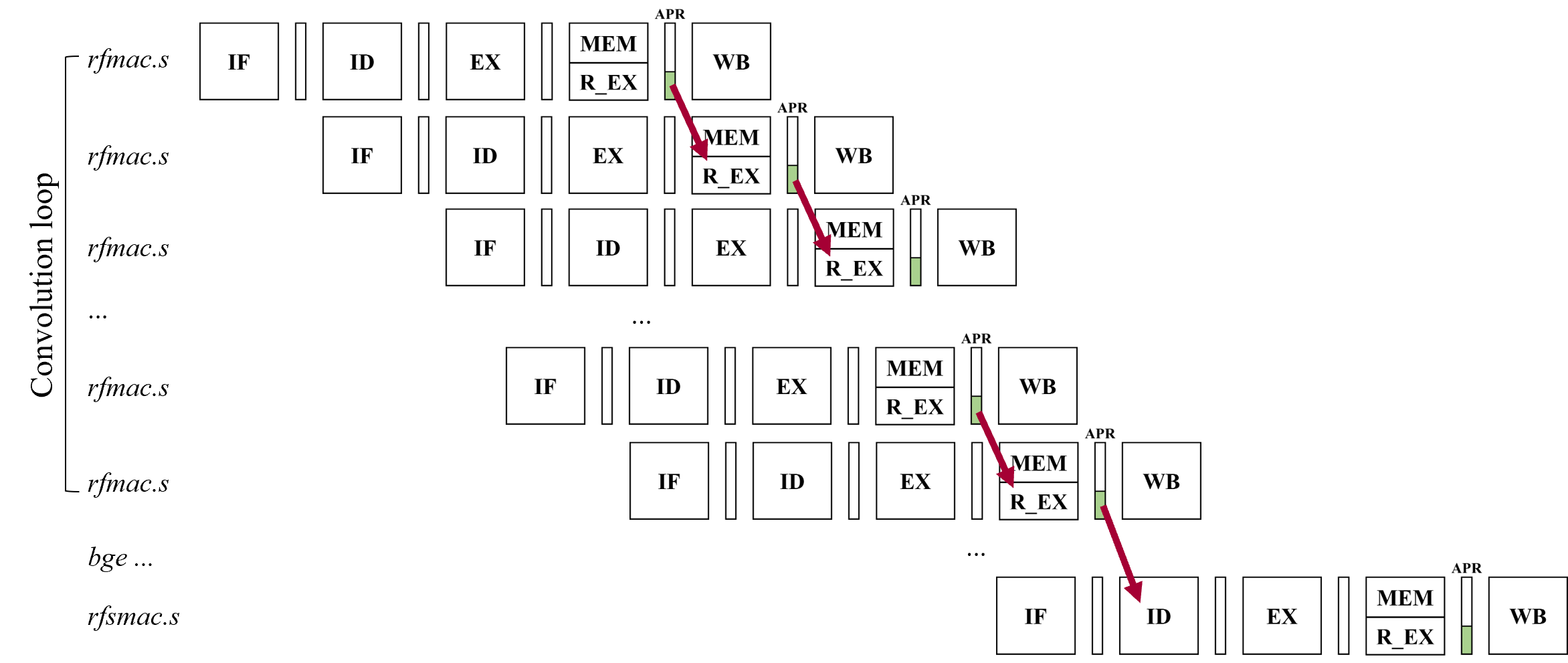}
    \caption{Forwarding effect on MAC at R-extension.}
    \label{fig:forwarding}
\end{figure}

The use of APR in the R-extension enables efficient data forwarding for MAC operations described at \autoref{fig:forwarding}, reducing memory access. This approach contributes to DNN acceleration and power consumption reduction \cite{b11}, making the R-extension suitable for energy-sensitive applications in lightweight devices.

\subsection{RISC-V F Standard Extension}
The compiler utilizes the F-extension when computing convolution in original RISC-V architectures with FP parameters \cite{b1}. This extension handles single-precision FP convolution operations, primarily through \textit{fadd.s} (FP add) and \textit{fmul.s} (FP multiply) instructions shown at \autoref{fig:code}(a). These instructions are integral to the computational processes underlying DNNs.

\begin{table}[t!]
\centering
\caption{Format field encoding.}
\label{tab:fmt}
\begin{tabular}{|c|c|c|}
\hline
\rowcolor[HTML]{EFEFEF} 
\textbf{fmt field} & \textbf{Mnemonic} & \textbf{Meaning}        \\ \hline
00                 & S                 & 32-bit single-precision \\ \hline
01                 & D                 & 64-bit double-precision \\ \hline
10                 & H                 & 16-bit half-precision   \\ \hline
11                 & Q                 & 128-bit quad-precision  \\ \hline
\end{tabular}
\end{table}

As shown in \autoref{fig:isa}, in addition to requiring two types of source registers; a destination register and source register – which are standard for most RISC-V instructions – the FP operation instructions also necessitate fields to specify the format \textit{fmt} and rounding mode \textit{rm}. The format, encoded as a 2-bit field, indicates the precision of the FP. \autoref{tab:fmt} shows the format field encoding. In the case of the F-extension, this field is set to `00', corresponding to the mnemonic S for single-precision operations. Similarly, the D-extension, intended for double-precision, uses a different mnemonic, D.

This study introduces the R-extension, which also targets single-precision FP operations. It adopts the `00' format field convention from the F-extension, ensuring compatibility and facilitating integration into the existing RISC-V infrastructure. The rounding mode, crucial for the precision of computations, is defined in the control and status register (CSR), which guarantees adherence to the IEEE 754 standard \cite{b2} and maintains the necessary precision for the given application.

Our study proposes the R-extension, a novel approach tailored for DNN workloads building upon the capabilities of the F-extension. The R-extension introduces custom instructions, specifically \textit{rfmac.s} and \textit{rfsmac.s}, designed to reduce the instruction count (IC) and memory usage during DNN operations.

\subsection{Customized Instruction}
Various efforts have focused on optimizing DNN processing through specialized instructions within the RISC-V architecture's evolving landscape. One such example is the development of a Winograd-based convolution acceleration instruction \cite{b3}, which performs a convolution calculation between a \(4\times4\) input matrix and a \(4\times4\) convolution kernel matrix and generates a \(2\times2\) output matrix. While this custom instruction, \textit{conv23}, enhances the efficiency of matrix convolutions for specific convolutional neural network (CNN) algorithms, it is restricted in its applicability. It does not tackle the core issue of MAC operations that are heavily utilized not only in convolution and fully connected layers of CNNs but also in other DNN architectures. This emphasizes the requirement for more adaptable solutions catering to a wide range of DNN architectures and operations.

Further research introduces \textit{vmac} (MAC operation on RISC-V V-extension \cite{b1}) and \textit{vload} (vector register load) instructions, mainly through enhancements in dot product computations and data transfer efficiency \cite{b4}. By supporting single instruction, multiple data (SIMD) architecture, these extensions substantially enhance the computational performance of DNN tasks. However, \textit{vmac} is only for paralleling MAC operations; it does not improve memory usage. We transformed \textit{vmac} to work in an FP scalar process for an experiment, and the instruction is called \textit{fmac.s} with usage as \autoref{fig:code}(b), our baseline that implements a MAC module at the EX stage and is compared with the R-extension.

The R-extension represents not merely an optimization for scalar processor environments but also an expandable entity within the RISC-V architecture. Harmonizing with the established architectural framework offers a viable extension path toward SIMD V-extension integration. This versatility underscores the extension's potential to evolve with advancing vector processors, positioning the RISC-V architecture to meet the expanding computational demands.

\section{Methodology}

The specialized pipeline came out for MAC operation, and to support this architecture, there are two types of instructions.

\subsection{Instruction Set Architecture (ISA)}
To add custom instructions on RISC-V ISA to compile, we should first assign the instruction symbol, type, constituents, MASK, and MATCH. MASK is the 32-bit binary to filter out the opcode and functions(e.g., \textit{funt3}, \textit{funct5},\textit{ funct7}, etc). MATCH is the 32-bit binary that contains actual opcodes and functions for each designated seat on instruction. After selecting the components for compilation, we need to add the required format to \textit{riscv-gnu-toolchain} \cite{b5}, then reconfigure and build the compiler.

\begin{figure}[t!]
    \centering
    \includegraphics[width=1\linewidth]{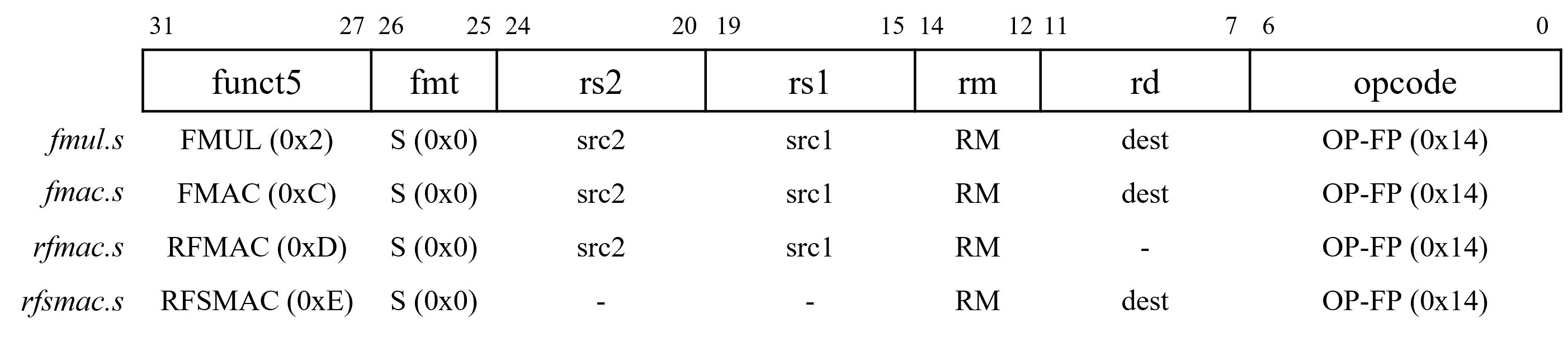}
    \caption{Instruction format of F-extension (\textit{fmul.s}), Baseline (\textit{fmac.s}), and R-extension (\textit{rfmac.s}, \textit{rfsmac.s}).}
    \label{fig:isa}
\end{figure}

\autoref{fig:isa} presents two innovative instructions within the RISC-V R-extension. The instruction \textit{rfmac.s} requires two source registers, \textit{rs1} and \textit{rs2}, and facilitates the multiplication of these sources, accumulating the result in the APR. This process eliminates the need for an immediate write-back to a destination register \textit{rd}, as the intermediate result is stored in the APR for further accumulation. As illustrated in \autoref{fig:code}(c-1), the \textit{rfsmac.s} instruction, employed at the end of the convolution loop where the final result is held in the APR, is tasked with writing this result from the APR to the \textit{rd} during the instruction-decode (ID) stage and resetting the APR at the MEM stage, as illustrated in \autoref{fig:dataflow}.

\begin{figure}[t!]
    \centering
    \includegraphics[width=1\linewidth]{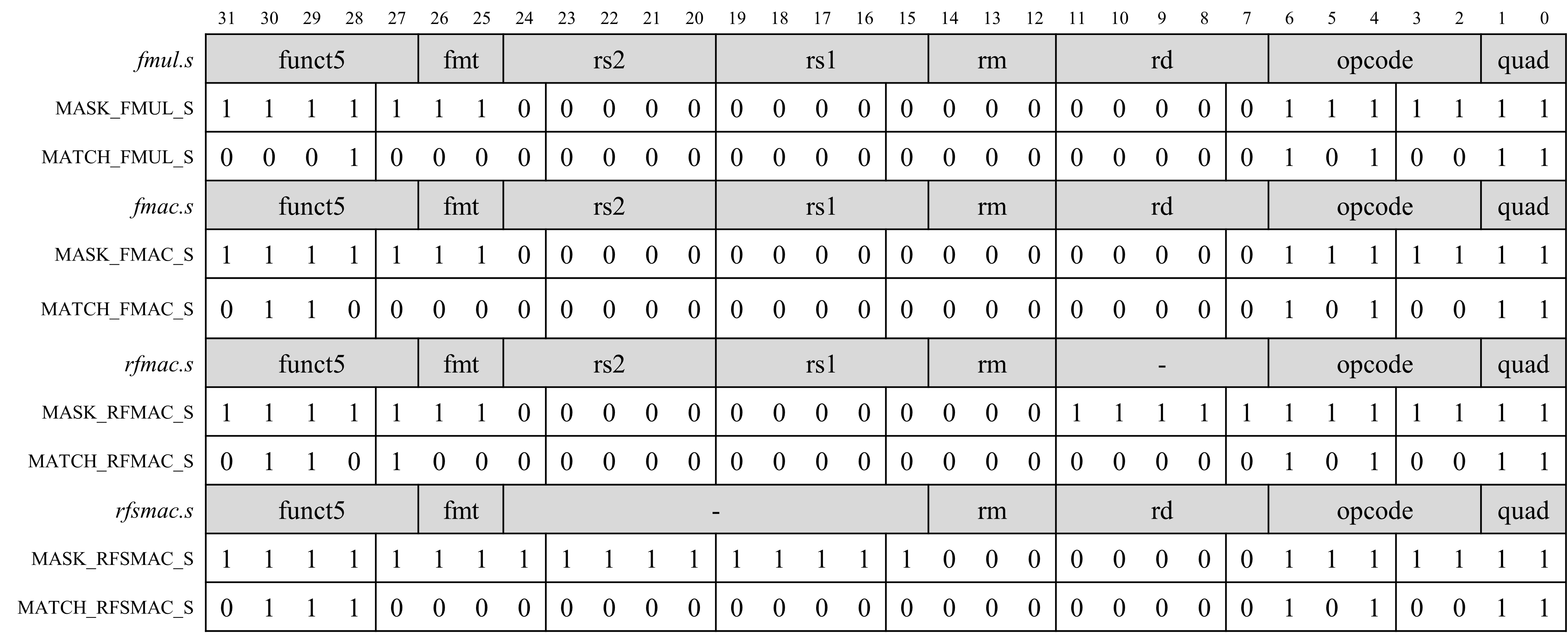}
    \caption{MASK and MATCH of F-extension, Baseline, and R-extension.}
    \label{fig:mask}
\end{figure}

The opcode for these instructions adheres to the F-extension to maintain consistency within FP operations. Unique functions mapped in \autoref{fig:mask} are assigned to each instruction to ensure no overlap with existing instructions. Newly set MASK and MATCH leave only the essential parts, such as \textit{rm}, \textit{rs1}, \textit{rs2}, and \textit{rd}.

\subsection{Microarchitecture}
\begin{figure}[htb!]
    \centering
    \includegraphics[width=1\linewidth]{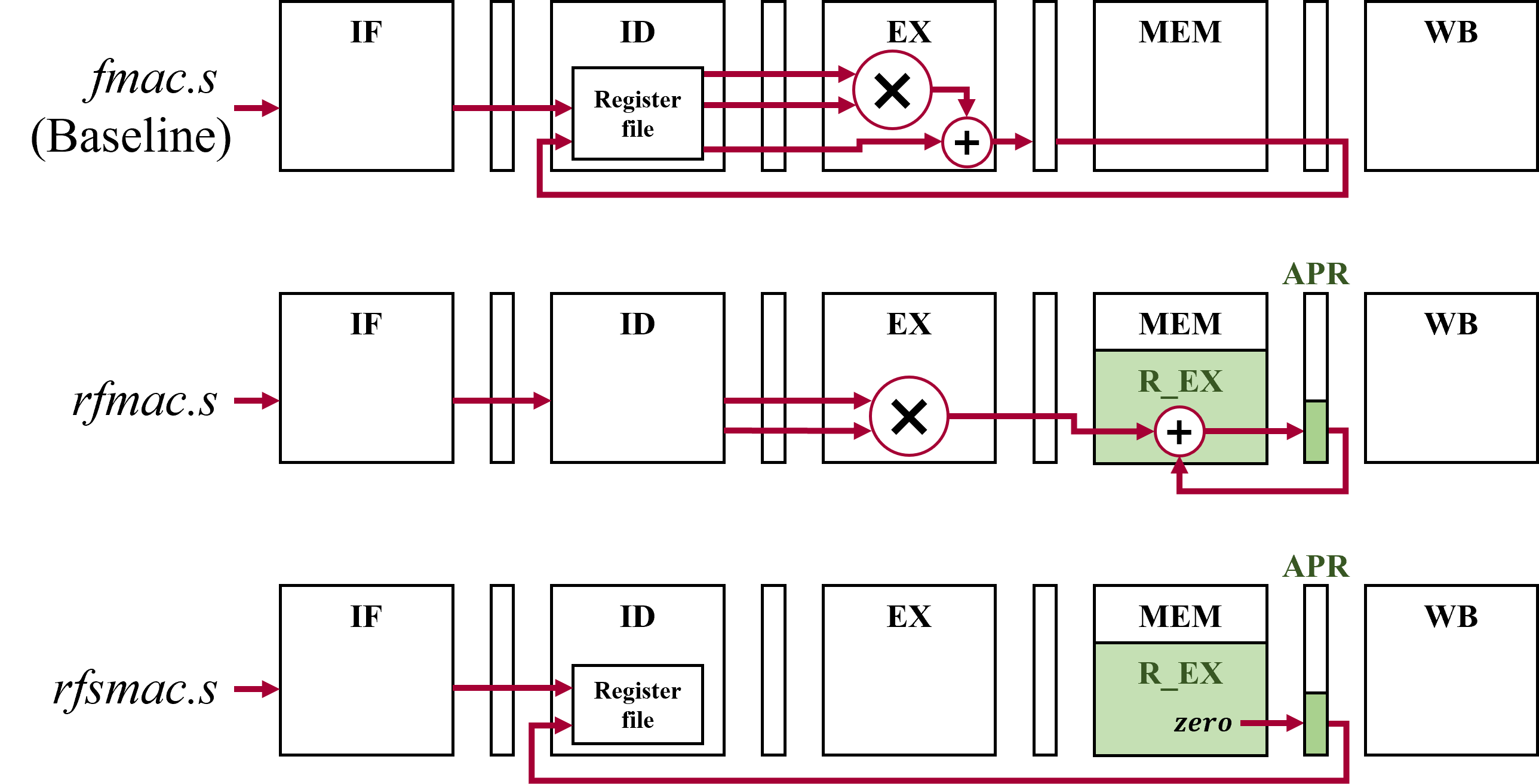}
    \caption{Dataflow of Baseline and R-extension with R\_EX and APR.}
    \label{fig:dataflow}
\end{figure}

The letter R on R-extension stands for the rented-pipelining concept. At the original pipeline stages, instruction-fetch (IF), ID, EX, MEM, and WB, each stage has its exclusive role. However, \textit{fmac.s} instruction of baseline in \autoref{fig:dataflow} does not use the MEM stage and just waits for one cycle, and data instantly goes to the WB stage. It is a waste of pipeline stage and area cost on DNN operation.

The rented-pipeline maintains the performance and number of pipeline stages even though it optimizes accumulation operation into the RISC-V CPU. In \autoref{fig:dataflow}, an extra accumulator in the MEM stage named the rented execution stage (R\_EX) is turned on instead of memory read and write when introduced instructions are fetched. Using EX and R\_EX stages, two of the five pipelines are used for execution without area dissipation.

APR is focused on reducing memory usage by eliminating unnecessary \textit{flw} (FP load) and \textit{fsw} (FP store) instructions during DNN operation as compared between \autoref{fig:code}(a-2), (b-2), and (c-2). In \autoref{fig:dataflow}, APR is located at the MEM/WB pipeline register.  It only needs an extra 32-bit register to keep updating the partial sum from the accumulator at the R\_EX stage. The input to the APR is determined by a MUX that selects between the accumulated data when the \textit{rfmac.s} instruction is executed and \textit{zero} for initialization when the \textit{rfsmac.s} instruction is issued. Furthermore, the output of APR is interfaced with the R\_EX stage to facilitate accumulation with newly multiplied data and with the ID stage, which is governed by the control bit determined by the presence of \textit{rfsmac.s} instruction. With these two hardware contributions, rented-pipeline and APR of R-extension, the CPU can accelerate DNN inference with low power and area overhead avoidance.

\begin{figure}[b!]
    \begin{lstlisting}[language=C, gobble=5, numbers=none]
    \#define rfmac_s(rs1, rs2) \
        asm volatile ("rfmac.s %0, %1" : : "f"(rs1), "f"(rs2))
    \#define rfsmac_s(rd) \
        asm volatile ("rfsmac.s %0" : "=f"(rd))
    \end{lstlisting}
    \caption{Definitions of inline assembly of C language. These definitions are used in \autoref{fig:code}.}
    \label{fig:define}
\end{figure}

\begin{table}[b!]
    \centering
    \caption{Configuration of simulation on gem5. All experiments are done in the same configuration.}
    \label{tab:config}
    \resizebox{\columnwidth}{!}{%
    \begin{tabular}{|
    >{\columncolor[HTML]{EFEFEF}}l |l|}
    \hline
    \textbf{CPU}                  & 1 GHz, Single-core, In-Order           \\ \hline
    \textbf{L1 Instruction Cache} & 512 KB, 2-way, 64 B line, 2 cycle latency \\ \hline
    \textbf{L1 Data Cache}        & 512 KB, 2-way, 64 B line, 2 cycle latency \\ \hline
    \textbf{Memory} & \begin{tabular}[c]{@{}l@{}}2 GB, DDR3, 1600 MT/s,\\8 banks, 8-bit wide data\end{tabular} \\ \hline
    \end{tabular}%
    }
\end{table}

\begin{table*}[t!]
\centering
\caption{Comparison of performance factors by LeNet, Resnet-20, MobileNet-V1(Scaled).}
\label{tab:result}
\arrayrulecolor{black}
\resizebox{\linewidth}{!}{%
\begin{tabular}{|l|l|r|r|r|r|r|} 
\hhline{~>{\arrayrulecolor{white}}->{\arrayrulecolor{black}}-----|}
\multicolumn{1}{c}{} & \multicolumn{1}{c|}{} & \multicolumn{1}{c|}{{\cellcolor[rgb]{0.902,0.902,0.902}}\textbf{runtime (Second)}} & \multicolumn{1}{c|}{{\cellcolor[rgb]{0.902,0.902,0.902}}\textbf{IC}} & \multicolumn{1}{c|}{{\cellcolor[rgb]{0.902,0.902,0.902}}\textbf{IPC}} & \multicolumn{1}{c|}{{\cellcolor[rgb]{0.902,0.902,0.902}}\textbf{memtype instructions}} & \multicolumn{1}{c|}{{\cellcolor[rgb]{0.902,0.902,0.902}}\textbf{L1 cache overall access}} \\ 
\hhline{|=======|}
{\cellcolor[rgb]{0.753,0.753,0.753}} & {\cellcolor[rgb]{0.902,0.902,0.902}}RV64F & 0.066 & 44,310,154 & 0.666 & 19,288,578 & 23,071,838 \\ 
\hhline{|>{\arrayrulecolor[rgb]{0.753,0.753,0.753}}->{\arrayrulecolor{black}}------|}
{\cellcolor[rgb]{0.753,0.753,0.753}} & {\cellcolor[rgb]{0.902,0.902,0.902}}Baseline & 0.048 & 35,792,547 & 0.740 & 16,043,778 & 19,841,884 \\ 
\hhline{|>{\arrayrulecolor[rgb]{0.753,0.753,0.753}}->{\arrayrulecolor{black}}------|}
{\cellcolor[rgb]{0.753,0.753,0.753}} & {\cellcolor[rgb]{0.902,0.902,0.902}}RV64R & 0.032 & 27,010,675 & 0.847 & 12,045,594 & 15,449,482 \\ 
\hhline{|>{\arrayrulecolor[rgb]{0.753,0.753,0.753}}->{\arrayrulecolor{black}}------|}
\rowcolor[rgb]{0.886,0.937,0.886} {\cellcolor[rgb]{0.753,0.753,0.753}} & {\cellcolor[rgb]{0.792,0.894,0.792}}Enhancement Over RV64F & 52.05 \% & 39.04 \% & 27.13 \% & 37.55 \% & 33.04 \% \\ 
\hhline{|>{\arrayrulecolor[rgb]{0.753,0.753,0.753}}->{\arrayrulecolor{black}}------|}
\rowcolor[rgb]{0.886,0.937,0.886} \multirow{-5}{*}{{\cellcolor[rgb]{0.753,0.753,0.753}}\textbf{LeNet }} & {\cellcolor[rgb]{0.792,0.894,0.792}}Enhancement Over Baseline & 34.05 \% & 24.54 \% & 14.43 \% & 24.92 \% & 22.14 \% \\ 
\hline
{\cellcolor[rgb]{0.753,0.753,0.753}} & {\cellcolor[rgb]{0.902,0.902,0.902}}RV64F & 6.210 & 4,103,496,569 & 0.661 & 1,795,154,166 & 2,103,847,934 \\ 
\hhline{|>{\arrayrulecolor[rgb]{0.753,0.753,0.753}}->{\arrayrulecolor{black}}------|}
{\cellcolor[rgb]{0.753,0.753,0.753}} & {\cellcolor[rgb]{0.902,0.902,0.902}}Baseline & 4.413 & 3,246,429,938 & 0.736 & 1,468,652,534 & 1,736,203,748 \\ 
\hhline{|>{\arrayrulecolor[rgb]{0.753,0.753,0.753}}->{\arrayrulecolor{black}}------|}
{\cellcolor[rgb]{0.753,0.753,0.753}} & {\cellcolor[rgb]{0.902,0.902,0.902}}RV64R & 2.691 & 2,352,965,745 & 0.874 & 1,062,330,923 & 1,289,180,424 \\ 
\hhline{|>{\arrayrulecolor[rgb]{0.753,0.753,0.753}}->{\arrayrulecolor{black}}------|}
\rowcolor[rgb]{0.886,0.937,0.886} {\cellcolor[rgb]{0.753,0.753,0.753}} & {\cellcolor[rgb]{0.792,0.894,0.792}}Enhancement Over RV64F & 56.66 \% & 42.66 \% & 32.30 \% & 40.82 \% & 38.72 \% \\ 
\hhline{|>{\arrayrulecolor[rgb]{0.753,0.753,0.753}}->{\arrayrulecolor{black}}------|}
\rowcolor[rgb]{0.886,0.937,0.886} \multirow{-5}{*}{{\cellcolor[rgb]{0.753,0.753,0.753}}\textbf{ResNet20 }} & {\cellcolor[rgb]{0.792,0.894,0.792}}Enhancement Over Baseline & 39.02 \% & 27.52 \% & 18.85 \% & 27.67 \% & 25.75 \% \\ 
\hline
{\cellcolor[rgb]{0.753,0.753,0.753}} & {\cellcolor[rgb]{0.902,0.902,0.902}}RV64F & 7.035 & 4,923,965,486 & 0.700 & 2,130,037,330 & 2,599,414,994 \\ 
\hhline{|>{\arrayrulecolor[rgb]{0.753,0.753,0.753}}->{\arrayrulecolor{black}}------|}
{\cellcolor[rgb]{0.753,0.753,0.753}} & {\cellcolor[rgb]{0.902,0.902,0.902}}Baseline & 5.255 & 4,122,177,959 & 0.784 & 1,824,588,370 & 2,222,467,107 \\ 
\hhline{|>{\arrayrulecolor[rgb]{0.753,0.753,0.753}}->{\arrayrulecolor{black}}------|}
{\cellcolor[rgb]{0.753,0.753,0.753}} & {\cellcolor[rgb]{0.902,0.902,0.902}}RV64R & 3.720 & 3,307,689,859 & 0.889 & 1,453,124,800 & 1,813,851,904 \\ 
\hhline{|>{\arrayrulecolor[rgb]{0.753,0.753,0.753}}->{\arrayrulecolor{black}}------|}
\rowcolor[rgb]{0.886,0.937,0.886} {\cellcolor[rgb]{0.753,0.753,0.753}} & {\cellcolor[rgb]{0.792,0.894,0.792}}Enhancement Over RV64F & 47.12 \% & 32.82 \% & 27.04 \% & 31.78 \% & 30.22 \% \\ 
\hhline{|>{\arrayrulecolor[rgb]{0.753,0.753,0.753}}->{\arrayrulecolor{black}}------|}
\rowcolor[rgb]{0.886,0.937,0.886} \multirow{-5}{*}{{\cellcolor[rgb]{0.753,0.753,0.753}}\textbf{MobileNet V1 }} & {\cellcolor[rgb]{0.792,0.894,0.792}}Enhancement Over Baseline & 29.21 \% & 19.76 \% & 13.34 \% & 20.36 \% & 18.39 \% \\ 
\hline
\rowcolor[rgb]{1,0.91,0.91} {\cellcolor[rgb]{0.753,0.753,0.753}} & {\cellcolor[rgb]{1,0.792,0.792}}Enhancement Over RV64F & 51.94 \% & 38.18 \% & 28.82 \% & 36.72 \% & 33.99 \% \\ 
\hhline{|>{\arrayrulecolor[rgb]{0.753,0.753,0.753}}->{\arrayrulecolor{black}}------|}
\rowcolor[rgb]{1,0.91,0.91} \multirow{-2}{*}{{\cellcolor[rgb]{0.753,0.753,0.753}}\textbf{Overall }} & {\cellcolor[rgb]{1,0.792,0.792}}Enhancement Over Baseline & 34.09 \% & 23.94 \% & 15.54 \% & 24.32 \% & 22.09 \% \\
\hline
\end{tabular}
}
\end{table*}

\section{Experimental Results}
\subsection{Simulation Result}
To use R-extension instructions in C language, we should use inline assembly `asm' and `volatile' like \autoref{fig:define}. After compiling with customized \textit{riscv-gnu-toolchain}, the binary file containing R-extension instructions came out. Highlights on the right side of \autoref{fig:code} are the main instructions that are more repeated than others. F-extension has six instructions with four memory loads, one memory store, and two arithmetic operations. Baseline, with little advanced, has five instructions with three memory loads, one memory store, and one operation. R-extension reduces half of the memory-related instructions compared with the previous two architecture and arithmetic operation instructions. This condition leads to DNN acceleration with low area cost and power consumption described below.

Before using the gem5 simulator \cite{b6}, we define the processor execution behavior in the gem5 instruction decoding algorithm. We make gem5 to accumulate data into the internal register APR when the opcode and function of \textit{rfmac.s} are detected. In the case of \textit{rfsmac.s}, the APR data is stored in the destination register and reset to \textit{zero}.

As illustrated in \autoref{tab:config}, we configured a system with separate L1 instruction and data caches, each with specific parameters such as size, associativity, and latency. The total cache size is set to 1 MB,  operating under a 1 GHz clock domain. This setting was chosen to align with ARM Cortex-A55 specifications \cite{b13}, which is usually used for home appliances SoC, such as smart TVs, representing the most demanding DNN inference scenarios. Such alignment ensures that our simulation environment realistically reflects the performance requirements of these applications. This setup utilizes a one-level cache architecture to precisely monitor the CPU's memory access patterns via the L1 cache, providing insights for understanding overall system performance.

We benchmark DNN inference on edge devices using three models: LeNet for simple tasks, ResNet-20 for moderate complexity, and MobileNet-V1 for advanced applications, assessing a range from basic to complex AI capabilities \cite{b7, b8, b9}. RV64F is a RISC-V 64-bit F-extension that contains FP multiplication and addition, as well as FP load and storage. The baseline architecture is RV64F with a na\"ive MAC operation module integrated into the EX stage, and \textit{fmac.s} instruction activates the module. We evaluate five major results: simulation runtime, IC, IPC, number of instructions for memory operations, and number of L1 cache accesses.

\begin{table}[h!]
\centering
\caption{Implementation result of Baseline and RV64R with overhead.}
\label{tab:imp_result}
\begin{tabular}{|l|r|r|r|} 
\hhline{~---|}
\multicolumn{1}{l|}{} & \multicolumn{1}{l|}{{\cellcolor[rgb]{0.937,0.937,0.937}}\textbf{Baseline}} & \multicolumn{1}{l|}{{\cellcolor[rgb]{0.937,0.937,0.937}}\textbf{RV32R}} & \multicolumn{1}{l|}{{\cellcolor[rgb]{0.937,0.937,0.937}}\textbf{Overhead}} \\ 
\hhline{|====|}
{\cellcolor[rgb]{0.937,0.937,0.937}}\textbf{LUT} & 1587 & 1559 & -1.76 \% \\ 
\hline
{\cellcolor[rgb]{0.937,0.937,0.937}}\textbf{FF} & 1965 & 1997 & 1.63 \% \\ 
\hline
{\cellcolor[rgb]{0.937,0.937,0.937}}\textbf{I/O} & 357 & 357 & 0 \% \\
\hline
\end{tabular}
\end{table}
\autoref{tab:result} shows that the enhancement rates are expressed in percentages, highlighting the performance gains of RV64R over RV64F and baseline. Across all models, RV64R demonstrates substantial enhancements in all results compared to RV64F and baseline. This indicates that the RV64R architecture is more efficient and faster than RV64F and the baseline configuration.

The overall advancement rates compile the gains across all three neural network models, providing a comprehensive view of the performance boost. The RV64R architecture has advanced by about 50\% over RV64F and about 32\% over the baseline in terms of runtime, marking a significant increase in speed. The progress is also noteworthy for other metrics, especially regarding IPC and L1 cache accesses, which are vital for rapid and efficient processing.

\subsection{Implementation Result}
To explore the resource overhead of our R-extension, we use the Xilinx Vivado implementation 2023 tool to synthesize and implement the R-extension core and baseline core; F-extension with Na\"ive MAC operation into the xcvu095-ffva2104-2-e FPGA device. The Vivado tool recommends synthesizing FP IP (FP multiplier, FP adder) into DSP modules. However, we changed the options to synthesize them into LUT for comparative clarity. The resource utilization results are shown in \autoref{tab:imp_result}. Compared with the baseline, the R-extension core overhead is 1.76\% less LUT and 1.63\% more FF resources. This result is primarily due to the R-extension's design, which involves adding only a few multiplexers (MUXs) and altering the position of the accumulator, thus minimizing changes to the overall architecture. It is a tiny cost effect on changing the F-extension into the proposed R-extension, and it is worth considering the improvements in the upper result.

\section{Conclusion}

The research presented introduces advancements in RISC-V architecture through R-extension (RV64R). This extension, which incorporates novel elements like the rented-pipeline and APR, enhances the efficiency of MAC operations, a critical aspect for processing in neural network models. By integrating the new instructions \textit{rfmac.s} and \textit{rfsmac.s}, the RV64R architecture demonstrates performance improvements over the F-extension and the baseline architecture in neural network inference. This is evident in metrics such as benchmark runtime, IPC, and L1 cache accesses, where R-extension outperforms its counterparts, indicating a more efficient and faster processing capability. Notably, reducing cache accesses also leads to power consumption savings, a critical factor in designing energy-efficient computing systems \cite{b14}.

Moreover, the research delves into the practical aspects of implementing the R-extension. Utilizing the Xilinx Vivado tool for RV64R synthesis and implementation, the study highlights the minimal resource overhead in transitioning from the baseline to the proposed R-extension. The slight increase in FF resources and the decrease in LUT usage underline the feasibility and efficiency of the R-extension. Considering the substantial performance gains in processing speed and efficiency, particularly in neural network models, and the added advantage of reduced power consumption through decreased cache access, the RV64R emerges as a valuable contribution to the RISC-V ecosystem, paving the way for efficient computing solutions in various applications.

\section*{Acknowledgment}
This work was partly supported by Institute of Information \& communications Technology Planning \& Evaluation(IITP) grant funded by the Korea government(MSIT) (No.2019-0-00421, AI Graduate School Support Program(Sungkyunkwan University)); in part by the Competency Development Program for Industry Specialists of the Korean Ministry of Trade, Industry and Energy (MOTIE), operated by Korea Institute for Advancement of Technology (KIAT) (No. P0023704, Semiconductor-Track Graduate School(SKKU)); in part by the Ministry of Trade, Industry and Energy (MOTIE) under Grant 20011074; in part by the Technology Innovation Program (or Industrial Strategic Technology Development Program-Public-Private Joint Investment Advanced Semiconductor Talent Development Project) (RS-2023-00237136, Development of CXL/DDR5-based memory subsystem for AI accelerators) funded By the Ministry of Trade, Industry \& Energy(MOTIE, Korea)(1415187686)

\end{document}